\documentclass[twocolumn,
showpacs,preprintnumbers,
showkeys,
amsmath,amssymb,
prl,reprint
]{revtex4-1}

\usepackage{graphicx}
\usepackage{dcolumn}
\usepackage{bm}
\usepackage{hyperref}

\begin{document}


\title{Controlling spin relaxation in hexagonal BN-encapsulated graphene with a transverse electric field}

\author{M. H. D. Guimar\~aes}
 \email{m.h.diniz.guimaraes@rug.nl}
\author{P. J. Zomer}
\author{J. Ingla-Ayn\'es}
\author{J. C. Brant}
\author{N. Tombros}
\author{B. J. van Wees}
\affiliation{Physics of Nanodevices, Zernike Institute for Advanced Materials, University of Groningen, The Netherlands}

\date{\today}

\begin{abstract}
We experimentally study the electronic spin transport in hBN encapsulated single layer graphene nonlocal spin valves.
The use of top and bottom gates allows us to control the carrier density and the electric field independently.
The spin relaxation times in our devices range up to 2 ns with spin relaxation lengths exceeding 12 $\mu$m even at room temperature.
We obtain that the ratio of the spin relaxation time for spins pointing out-of-plane to spins in-plane is $\tau_{\bot} / \tau_{||} \approx$ 0.75 for zero applied perpendicular electric field.
By tuning the electric field this anisotropy changes to $\approx$0.65 at 0.7 V/nm, in agreement with an electric field tunable in-plane Rashba spin-orbit coupling.
\end{abstract}

\maketitle

The generation, manipulation and detection of spin information has been the target of several studies due to the implications for novel spintronic devices \cite{WolfScience2001,ZuticActaPhysicaSlovaca2007}.
In the recent years graphene has attracted a lot of attention in spintronics due to its theoretically large intrinsic spin relaxation time and length of the order of $\tau_{s}\approx$ 100 ns and $\lambda_{s}\approx$ 100 $\mu$m respectively \cite{BrataasPhys.Rev.Lett.2009,BarnafmmodePhys.Rev.B2011}.
Although experimental results still fall short of these expectations \cite{MihaiPhys.Rev.B2009,KawakamiPhys.Rev.Lett.2011,OezyilmazPhys.Rev.Lett.2011,GuimaraesNanoLetters2012}, graphene has already achieved the longest measured nonlocal spin relaxation length \cite{KawakamiPhys.Rev.Lett.2011,WojtaszekPhys.Rev.B2013} and furthest transport of spin information at room temperature \cite{ZomerPhys.Rev.B2012}.
However, the mechanisms for spin relaxation in graphene are still under heavy debate with various theoretical models proposed \cite{BrataasPhys.Rev.Lett.2009,BarnafmmodePhys.Rev.B2011,FabianPhys.Rev.Lett.2014,WuNewJournalofPhysics2012,RocheJPD2014,GuineaPhys.Rev.Lett.2012}.

To take advantage of the long spin relaxation times in graphene, e.g. for spin logic devices, one requires easy control of the spin information, for example by an applied electric field.
Single layer graphene is an ideal system for this purpose, not only because of its high mobilities and low intrinsic spin-orbit fields (SOF), but also due to the simple relation between the carriers' wavevector, the applied perpendicular electric field and the induced Rashba SOF \cite{BarnafmmodePhys.Rev.B2011,FabianPhys.Rev.B2009,BrataasPhys.Rev.Lett.2009,BrataasPhys.Rev.B2006,MelePhys.Rev.Lett.2005,MacDonaldPhys.Rev.B2006,RashbaPhys.Rev.B2009,FabianPhys.Rev.B2009a}.
In bilayer graphene a more complicated behavior is expected when spin-orbit coupling is considered \cite{FabianPhys.Rev.B2012}.

Here we report nonlocal spin transport measurements on single layer graphene in which we address both topics specified above.
Our devices consist of a single layer graphene flake on hexagonal Boron Nitride (hBN) of which a central region is encapsulated with another hBN flake and hence protected from the environment.
The presence of a top and bottom gate give rise to two independent electric fields that are experienced by the graphene: $E_{tg}= - \epsilon_{tg}(V_{tg}-V_{tg}^{0})/d_{tg}$ and $E_{bg}=\epsilon_{bg}(V_{bg}-V_{bg}^{0})/d_{bg}$, respectively \cite{WangNature2009}, where $\epsilon_{tg(bg)}\approx$ 3.9 is the dielectric constant, $d_{tg(bg)}$ is the dielectric thickness and $V_{tg(bg)}^{0}$ the position of the charge neutrality point for the top (bottom) gate.
Their difference controls the carrier density in the graphene ($n = (E_{bg} - E_{tg}) \epsilon_{0} / e$) and their average gives the effective electric field experienced by the graphene ($\bar{E}=(E_{tg}+E_{bg})/2$), which breaks the inversion symmetry in the encapsulated region, where $\epsilon_{0}$ is the electric constant and $e$ the electric charge.
Our devices show enhanced spin relaxation times of at least 2 ns and also, due to the higher electronic mobility, spin relaxation lengths above 12 $\mu$m at room temperature (RT) and 4.2 K.
By a simple model we show that the measured spin relaxation times are a lower bound due to the influence of the non-encapsulated regions.

By comparing the spin relaxation time for spins out-of-plane ($\tau_{\bot}$) to spins in-plane ($\tau_{||}$) as a function of the electric field we get insight on the nature of the SOF that cause spin relaxation in graphene.
For SOF pointing preferentially in the graphene plane, e.g. for adatoms and impurities, we expect: $\tau_{\bot}\approx 0.5\tau_{||}$ \cite{ZuticActaPhysicaSlovaca2007,BrataasPhys.Rev.Lett.2009,TombrosPhys.Rev.Lett.2008,FabianPhys.Rev.B2009a}.
If the SOF point out-of-plane, as for ripples\cite{BrataasPhys.Rev.Lett.2009}, we have: $\tau_{\bot} \gg \tau_{||}$.
However, if the main relaxation mechanism is through random magnetic impurities or defects, no preferential direction for the spins is expected: $\tau_{||} \approx \tau_{\bot}$.
Here we obtain $\tau_{\bot} / \tau_{||}\approx$ 0.75 at $\bar{E}$ = 0 V/nm$^{-1}$.
This ratio decreases with increasing $\bar{E}$, in agreement with an electric field induced Rashba SOF pointing in the graphene plane.

\begin{figure}[h]
	\centering
		\includegraphics[width=0.5\textwidth]{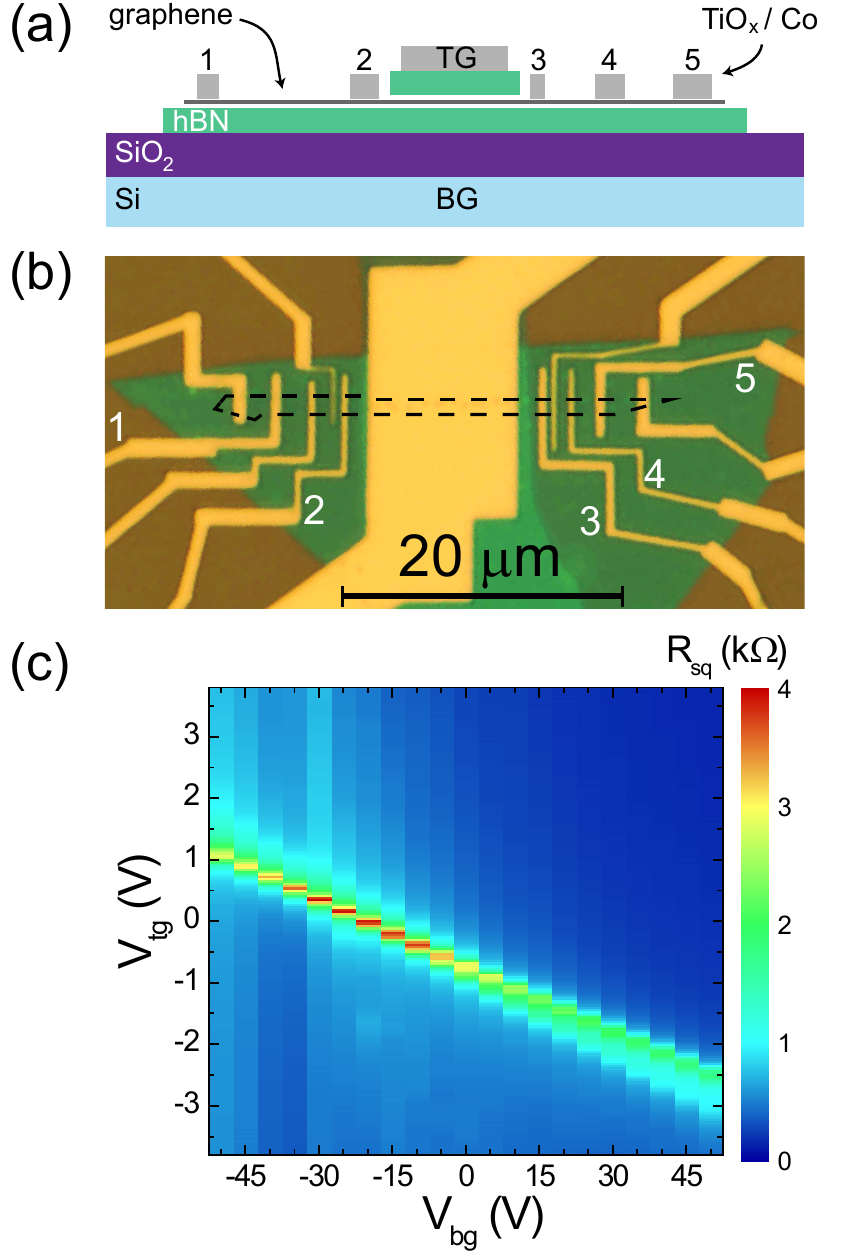}
	\caption{(a) Side-view schematics and (b) Top-view optical microscope image of a hBN encapsulated graphene spin-valve. The numbers show the contact electrodes and the top (bottom) gates electrodes are indicated as TG (BG). The graphene is outlined by the dashed line. (c) Square resistance (R$_{sq}$) as a function of $V_{tg}$ and $V_{bg}$.}
	\label{fig:figure-1}
\end{figure}

Device number 1 is illustrated in Fig. \ref{fig:figure-1}a and b.
The hBN-graphene-hBN stack sits on a 300 nm thick SiO$_{2}$ layer on a heavily doped Si substrate which is used as a back-gate.
The sample preparation is described in detail in the supplementary information and follows Ref. \cite{DeanScience2013,WeesAppliedPhysicsLetters2014}.
We use Co electrodes with a thin TiO$_{2}$ interface barrier to perform spin transport measurements.
Three devices were studied, all showing similar results.
Here we show the results for the device with the longest encapsulated region ($\approx$ 12 $\mu$m) and spacing between the inner contacts (13.8 $\mu$m).

The charge transport properties of the encapsulated region are measured by applying a current between electrodes 1 and 5, Fig. \ref{fig:figure-1}a and b, and scanning the top and bottom gate voltages ($V_{tg}$ and $V_{bg}$, respectively) while recording the voltage between electrodes 2 and 3.
Fig. \ref{fig:figure-1}c shows the square resistance ($R_{sq}$) as a function of $V_{tg}$ and $V_{bg}$.
The charge neutrality point depends on both $V_{tg}$ and $V_{bg}$ in a linear fashion.
The slope of the line gives the ratio between the bottom and top gate capacitances: $\alpha_{bg}/\alpha_{tg}\approx$ 0.036.
A small top gate independent resistance peak around $V_{bg}$ = -16.6 V (not visible in Fig. \ref{fig:figure-1}c) arises from the non-top gated regions between the two inner contacts.
The electronic mobility for this device is $\mu\approx$ 1.5 m$^{2}$/Vs at RT and $\mu\approx$ 2.3 m$^{2}$/Vs at 4.2 K.
Although the mobilities of our devices are above the best devices based on SiO$_{2}$ they are still one order of magnitude lower than the best devices on hBN \cite{DeanScience2013} which can be attributed to small bubbles or contamination visible on the graphene/hBN stack.

Spin dependent measurements are performed using a standard nonlocal geometry in which the current path is separated from the voltage detection circuit \cite{MihaiPhys.Rev.B2009}.
The current is driven between electrodes 1 and 2 and the voltage measured between electrodes 3 and 5, which are on the other side of the encapsulated region (Fig. \ref{fig:figure-1}a and b).
To obtain the spin relaxation time ($\tau_{s}$) and the spin diffusion coefficient ($D_{s}$) we perform Hanle precession measurements where the nonlocal signal is measured as a function of a perpendicular magnetic field $B$.
We then fit the data with the solution to the Bloch equations \cite{MihaiPhys.Rev.B2009}.

\begin{figure}[h]
	\centering
		\includegraphics[width=0.5\textwidth]{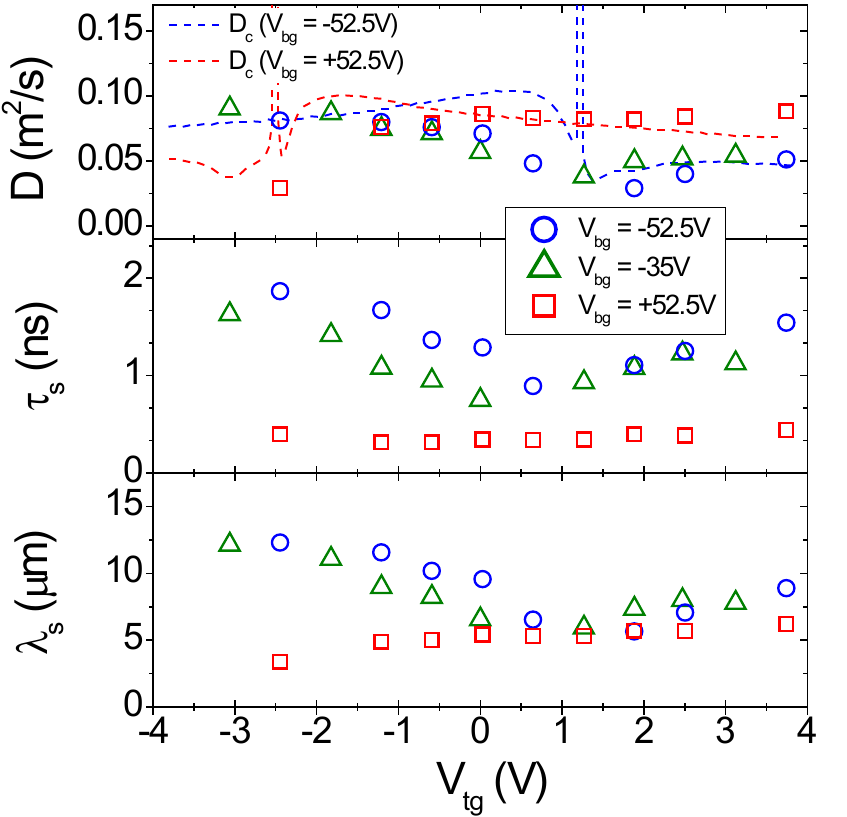}
	\caption{(Spin) diffusion coefficient ($D_{s}$), spin relaxation time $\tau_{s}$ and spin relaxation length $\lambda_{s}$ as a function of $V_{tg}$ for three different values of $V_{bg}$. The error bars are smaller than the dot size. The dashed red (blue) lines show the charge diffusion coefficient $D_{c}$ for $V_{bg}$ = +52.5 V (-52.5 V).}
	\label{fig:figure-2}
\end{figure}

The results for $D_{s}$, $\tau_{s}$ and the spin relaxation length ($\lambda_{s}=\sqrt{D_{s}\tau_{s}}$) as a function of the $V_{tg}$ for three values of $V_{bg}$ at 4.2 K are shown in Fig. \ref{fig:figure-2}.
A similar set of measurements was performed at RT and for other samples where only a small difference was observed (see supplementary information \cite{supinfo}).

Due to our device mobility, $D_{s}$ is higher than for regular graphene devices on SiO$_{2}$ ($D_{s}$ $\approx$ 0.02 m$^{2}$/s) \cite{KawakamiPhys.Rev.Lett.2011,MihaiPhys.Rev.B2009} and comparable to suspended \cite{GuimaraesNanoLetters2012} and non-encapsulated hBN supported devices \cite{ZomerPhys.Rev.B2012} ($D_{s}$ $\approx$ 0.05 m$^{2}$/s).
As an extra confirmation, we check that $D_{s}$ agrees with the charge diffusion coefficient $D_{c}=[R_{sq}e^{2}\nu(E_{F})]^{-1}$\footnote{A resistance of 3.2 k$\Omega$ was subtracted in the calculation of $D_{c}$ for $V_{bg}=-52.5$V to account for the non-top gated regions.}, where $e$ is the electron charge and $\nu(E_{F})$ the density of states at the Fermi energy $E_{F}$.
Next, we observe that the obtained spin relaxation times are higher than those on regular SiO$_{2}$ substrates ($\tau_{s}\approx$ 0.1 - 1 ns)\cite{KawakamiPhys.Rev.Lett.2011,MihaiPhys.Rev.B2009} and in non-encapsulated hBN supported devices ($\tau_{s}\approx$ 0.1 - 0.5 ns)  \cite{ZomerPhys.Rev.B2012}, reaching up to $\tau_{s}$=(1.9$\pm$0.2) ns at 4.2 K and $\tau_{s}$=(2.4$\pm$0.4) ns for RT.
These values surpass all previous nonlocal measurements of $\tau_{s}$ in single layer graphene both at room and low temperatures \footnote{2-terminal local measurements on epitaxial graphene estimated $\tau_{s}\approx$ 100 ns at 4.2 K \cite{FertNatPhys2012}.} \cite{FertNatPhys2012}.
We obtain a maximum of $\lambda_{s}$ = 12.3 $\mu$m at 4.2 K and $\lambda_{s}$ = 12.1 $\mu$m at RT.


Comparing $\tau_{s}$ obtained in our devices with non-encapsulated hBN based devices ($\tau_{s}\approx$ 0.2 ns) \cite{ZomerPhys.Rev.B2012}, we can conclude that the encapsulation of graphene on hBN significantly increases the spin relaxation times.
Note that the non-encapsulated devices had comparable electronic mobilities which indicates that $\tau_{s}$ is not linked to the momentum relaxation time in a trivial manner \cite{K.NanoLetters2012}.
By measuring a region about 5 $\mu$m away from the encapsulated part, we find $D_{s}\approx$ 0.03 m$^{2}$/s, $\tau_{s}\approx$ 0.3 ns and $\lambda_{s}\approx$ 3 $\mu$m, in agreement with the previously reported results.
The increase in $\tau_{s}$ for the encapsulated region can be due to several factors.
This region is protected from polymer remains or other contamination which can increase spin scattering.
In addition to that, the inversion asymmetry, which can generate an extra term for the spin-orbit coupling, is also reduced and controlled by tuning $V_{tg}$ and $V_{bg}$ separately as explained earlier.

As can be seen in Fig. \ref{fig:figure-2}, $\tau_{s}$ is modulated by $V_{tg}$, showing a dip close to the charge neutrality point in the encapsulated region (e.g. $V_{tg}$=1.1 V and $V_{bg}$=-52.5 V).
For larger charge carrier densities in the non-encapsulated regions, the modulation in $\tau_{s}$ by $V_{tg}$ is smaller, although still present.
Furthermore, the average value of $\tau_{s}$ is maximum at low carrier densities in the non-encapsulated regions (large negative $V_{bg}$) and decreases with increasing $V_{bg}$.
This difference on the measured $\tau_{s}$ for different carrier densities on the outer regions can be explained by simulations that treat the full device \cite{GuimaraesNanoLetters2012}.
Since the transport is diffusive, the spins can explore both the encapsulated and the non-encapsulated regions before been detected.
We take this into account by describing our sample as two outer regions connected by a central region \cite{supinfo}.
The relevant parameters ($R_{sq}$, $D_{s}$ and $\tau_{s}$) are set for each region individually.
The values for the outer regions and $R_{sq}$ and $D_{s}$ for the inner region can be extracted from our charge and spin transport measurements.

\begin{figure}[h]
	\centering
		\includegraphics[width=0.5\textwidth]{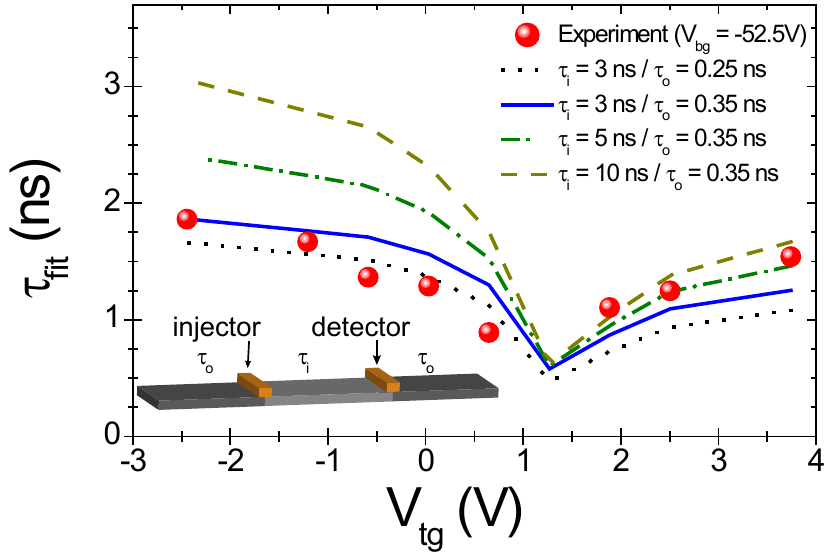}
	\caption{Effective spin relaxation times extracted for different values for $\tau_{o}$ and $\tau_{i}$ (lines) compared to our experimental data for $V_{bg}$ = -52.5 V (dots). The inset shows a schematics of the simulated system.}
	\label{fig:figure-3}
\end{figure}

Changing the values of the spin relaxation time for the outer regions, $\tau_{o}$, around the experimentally obtained values and changing the values for the spin relaxation time in the inner region, $\tau_{i}$, we simulate Hanle precession curves that are fitted in the same way done for our experiments to obtain an effective value for the spin relaxation time $\tau_{fit}$.
We get a reasonable quantitative agreement between our simulations and experiment at $V_{bg}$ = -52.5V for $\tau_{i}$ = 3 ns.
This means that the spin relaxation time for the encapsulated region is higher ($\tau_{s}\approx$ 3 ns), but still within the same order of magnitude as the values obtained by analyzing the data using a homogeneous system (Fig. \ref{fig:figure-2}).
The trend in $V_{tg}$ is also reproduced, which indicates that it is given by the ratio of the resistivities of the inner and outer regions.

Even though the experimentally obtained value for $\tau_{s}$ depends on the gate voltages in a non-trivial way due to the influence of the non-encapsulated regions, we can still study how the electric field affects the ratio between the spin relaxation times for out-of-plane to in-plane spins: $r=\tau_{\bot}/\tau_{||}$.
As explained in the introduction, this way we can get insight about the SOF in our system.

\begin{figure}[h]
	\centering
		\includegraphics[width=0.5\textwidth]{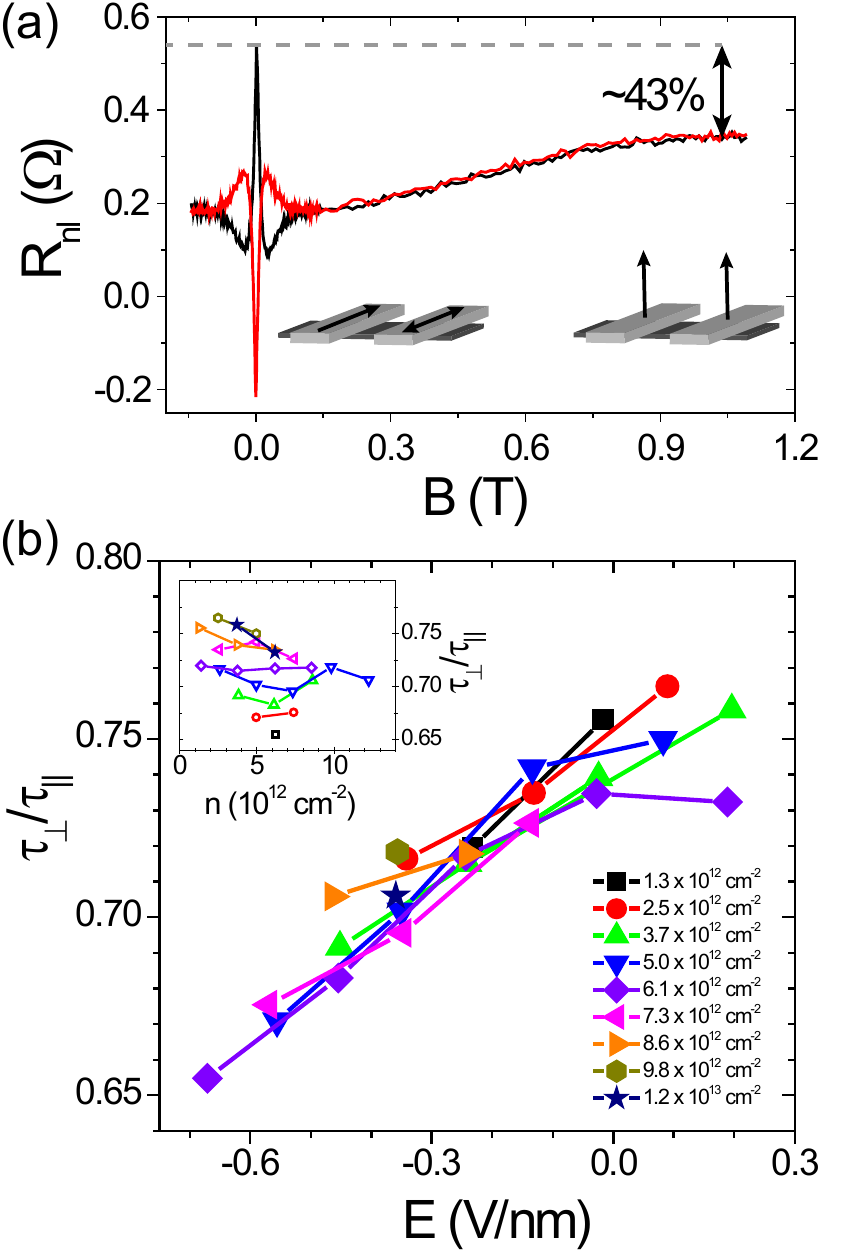}
	\caption{(a) $R_{nl}$ as a function of $B$ showing Hanle precession at low fields and a saturation of the signal at high fields when the magnetization of the electrodes point out-of-plane. Inset: A cartoon showing the magnetization of the contacts: in-plane at low fields and out-of-plane at high fields. (b) The ratio $\tau_{\bot} / \tau_{||}$ as a function of electric field $\bar{E}$ for different values of carrier density $n$. \textit{Inset}: The same data points for $\tau_{\bot} / \tau_{||}$ used in the main graph, but plotted as a function of $n$ where different colors and symbols represent the different values of $\bar{E}$.}
	\label{fig:figure-4}
\end{figure}

To compare the spin relaxation for spins parallel and perpendicular to the graphene plane we perform the Hanle precession measurements as described before, but increase the perpendicular magnetic field to higher values, $B>$ 1 T.
At such high magnetic fields the magnetization of the electrodes rotates out-of-plane and the injected spins do not precess anymore.
This is seen as a saturation of the nonlocal signal at high B, Fig. \ref{fig:figure-4}a.

We observe that different combinations of $V_{tg}$ and $V_{bg}$ result in a saturation of $R_{nl}$ at high magnetic fields at values always smaller than $R_{nl}$ at $B$=0 T, with the saturation occuring at 43-57$\%$ of the initial value.
Given that the nonlocal spin signal is given by $\Delta R_{nl}=\frac{P^{2}R_{sq}\lambda_{s}}{W}e^{-L/\lambda_{s}}$, where $P$ is the polarization of the electrodes, $L$ the distance between electrodes and $W$ the channel width, we can estimate the anisotropy in the spin relaxation times assuming that $P$ and $R_{sq}$ do not change significantly with field.
Due to a large magneto-resistance at low carrier densities, our analysis is done only for points at large enough carrier densities for both the inner and outer regions \footnote{At low $n$, $R_{sq}$ of our devices scales with B$^{2}$ leading to a background in our signal that overcomes the nonlocal spin signal at large B.}.
We can relate the ratio of the nonlocal spin signal and the ratio of the spin relaxation times by:
$R_{nl}^{\bot} / R_{nl}^{||} = \sqrt{r}e^{\frac{L}{\lambda_{||}}\left( \frac{\sqrt{r}-1}{\sqrt{r}} \right)}$, where $\lambda_{||}=\sqrt{D_{s}\tau_{||}}$ is obtained via our Hanle precession measurements.
In Fig. \ref{fig:figure-4}b we plot the ratio $\tau_{\bot} / \tau_{||}$ as a function of $\bar{E}$ for different values of $n$ where we see a clear decrease of this ratio, from 0.75 at $\bar{E}\approx$ 0 V/nm to about 0.65 at $\bar{E}\approx$  -0.7 V/nm.
The inset on Fig. \ref{fig:figure-4}b shows the dependence of $r$ as a function of $n$ for different values of $\bar{E}$ where no clear trend can be seen.
The value for $\tau_{\bot} / \tau_{||}$ for zero electric field is similar to the values found previously on SiO$_{2}$ based devices ($\tau_{\bot} \approx$ 0.8 $\tau_{||}$) \cite{TombrosPhys.Rev.Lett.2008}.
In the case of a inversion symmetric graphene layer with no extrinsic sources for SOF we would not expect any particular preference for direction of the spins, meaning that $\tau_{\bot} \approx \tau_{||}$ \cite{ZuticActaPhysicaSlovaca2007}.
The fact that even at $\bar{E}$ = 0 V/nm the ratio between $\tau_{\bot}$ and $\tau_{||}$ is below 1 means that even without an externally applied electric field there are probably remanent SOF pointing preferentially in the graphene plane.

The decrease of $r$ with increasing $\bar{E}$ is in agreement with theories that dictate an increase in in-plane Rashba SOF with the increase of electric field \cite{BarnafmmodePhys.Rev.B2011,FabianPhys.Rev.B2009,RashbaPhys.Rev.B2009,MacDonaldPhys.Rev.B2006,BrataasPhys.Rev.B2006,MelePhys.Rev.Lett.2005,FabianPhys.Rev.B2009a}.
The Rashba-type spin-orbit Hamiltonian for graphene is given by: $\mathcal{H}_{SO}=\frac{\Delta_{R}}{2} \left( \mathbf{\sigma} \times \mathbf{s} \right)_{z}$, where $\mathbf{\sigma}$ and $\mathbf{s}$ are the pseudospin and real spin Pauli matrices respectively.
For single layer graphene the spin-orbit constant is known to depend on the $\bar{E}$ in a linear way $\Delta_{R}=\zeta \bar{E}$ \cite{FabianPhys.Rev.B2009,RashbaPhys.Rev.B2009}, where $\zeta$ is the coupling constant.
Theoretical values range from $\zeta$ = 0.3 - 66 $\mu$eV/Vnm$^{-1}$ \cite{MacDonaldPhys.Rev.B2006,BrataasPhys.Rev.B2006,MelePhys.Rev.Lett.2005,FabianPhys.Rev.B2009}.
We can roughly estimate $\zeta$ by assuming a D'Yakonov-Perel mechanism for spin relaxation \cite{DP1971,FabianPhys.Rev.B2009a} with different values for the spin orbit coupling for in- and out-of-plane spins.
Our analysis \cite{supinfo} results in $\zeta \approx$ (40$\pm$20) $\mu$eV/Vnm$^{-1}$, within the range of the theoretical predictions.

In conclusion, we measured the spin transport characteristics of a single layer graphene device encapsulated with hBN.
We measured spin relaxation times up to $\tau_{s}\approx$ 2 ns and spin relaxation lengths above $\lambda_{s}$= 12 $\mu$m.
By taking into consideration that the non-encapsulated regions of our devices play a role in our measurements of $\tau_{s}$, we estimate the actual spin relaxation time in the encapsulated region to be $\tau_{s}\approx$ 3 ns.
Furthermore, we showed that the ratio between out-of-plane and in-plane spin relaxation times changes from $\tau_{\bot} / \tau_{||}\approx$ 0.75 to 0.65 with increasing the applied out-of-plane electric field.
This observation is in agreement with an electric field induced Rashba-type spin orbit.
Our results show not only that $\tau_{s}$ in graphene can be improved by improving the quality of the devices, but also that electrical control of spin information in graphene is possible, paving the way to new graphene spintronic devices.

\textit{Note added:} During the preparation of this manuscript we became aware of a work in which $\lambda_{s}$ up to 10 $\mu$m for single and few-layer graphene were achieved for a single gated structure which did not allow the study of $\tau_{s}$ as a function of $\bar{E}$ \cite{arxiv-aachen}.

We would like to acknowledge A. Kamerbeek and E. Sherman for insightful discussions and J. G. Holstein, H. M. de Roosz and H. Adema for the technical support.

The research leading to these results has received funding from the Dutch Foundation for Fundamental Research on Matter (FOM), the European Union Seventh Framework Programme under grant agreement n$^{\circ}$604391 Graphene Flagship, the People Programme (Marie Curie Actions) of the European Union's Seventh Framework Programme FP7/2007-2013/ under REA grant agreement n$^{\circ}$607904-13 Spinograph, NWO, NanoNed, the Zernike Institute for Advanced Materials and CNPq, Brazil.



%

\newpage
\clearpage


\section{Supplemental Material: Controlling spin relaxation in hexagonal BN-encapsulated graphene with a transverse electric field}


\maketitle

\subsection{Fabrication of the stacks}
The hBN-graphene-hBN stacks were fabricated using a pick-up method \cite{DeanScience2013} described elsewhere \cite{WeesAppliedPhysicsLetters2014} in more details.
The graphene and hBN flakes were obtained by mechanical cleavage of HOPG (SPI Supplies) and hBN powder (HQ Graphene).
The preparation of the stacks starts with the exfoliation and optical microscopy selection of hBN on a glass mask covered by a thin layer of polycarbonate (HQ Graphene).
The single layer graphene and bottom hBN flakes are exfoliated on Si/SiO$_{2}$ substrates and selected by optical microscopy.
The mask containing the top hBN flake is then aligned and pressed against the substrate containing the graphene flake and heated to $\approx$75 $^{\circ}$C.
When the mask is retracted from the substrate the graphene flake adhere strongly to the hBN flake and releases from the substrate sticking to the mask.
The mask containing the hBN-graphene stack is then aligned to the bottom hBN flake, brought in contact to the substrate and heated to temperatures up to $\approx$150 $^{\circ}$C.
At these temperatures the polycarbonate film melts on the substrate and releases from the mask together with the stack.
The sample is then left in chloroform for at least 15 hours in order to completely dissolve the polycarbonate film and obtain a clean graphene surface ensuring good graphene-contact interfaces.

\subsection{Contact deposition}
The contacts and top gate electrodes in this work were all fabricated at the same step to avoid further contamination of the non-encapsulated graphene regions.
The electrodes were patterned using standard e-beam lithography techniques using a PMMA 950K (270 nm thick) resist.
The markers for the e-beam lithography were patterned prior to the electrodes in the same PMMA resist and developed leaving openings which were used for the alignment of the electrodes.
We then used an e-beam evaporator with a base pressure below 10$^{-6}$ Torr to evaporate a 0.4 nm thick layer of Ti which was then naturally oxidized by inserting pure oxygen gas in the vacuum chamber to achieve pressures above 1 Torr.
After 15 minutes the chamber was pumped down once again to the initial base pressure and the procedure was repeated to deposit and oxidize another 0.4 nm thick Ti layer.
This procedure is then followed by the deposition of a 67 nm thick Co layer with a capping layer of Al (5 nm) to protect the Co layer from oxidizing.
The large thickness of the Co layer is necessary to both overcome the height of the bottom hBN flake and also to facilitate the alignment of the electrodes by the perpendicular magnetic field.
After the resist lift-off and wire bonding, the samples were loaded in a continuous He flow cryostat which stands in a room temperature electro-magnet.
The contact resistances were in the range 2 - 100 k$\Omega$.

\subsection{Transport measurements}
The charge transport measurements were performed by standard low-frequency lock-in techniques with currents up to 100 nA.
All spin transport measurements presented here were done using the non-local geometry in which the charge contribution to the measured signal is minimized by separating the charge current path from the voltage detection circuit.
We used standard low-frequency lock-in techniques with applied currents of 1 $\mu$A in order to preserve the high resistive contact interface barriers and avoid device heating. 

\subsection{Magnetoresistance effects on the nonlocal signal}
\begin{figure}[h]
	\centering
		\includegraphics[width=0.5\textwidth]{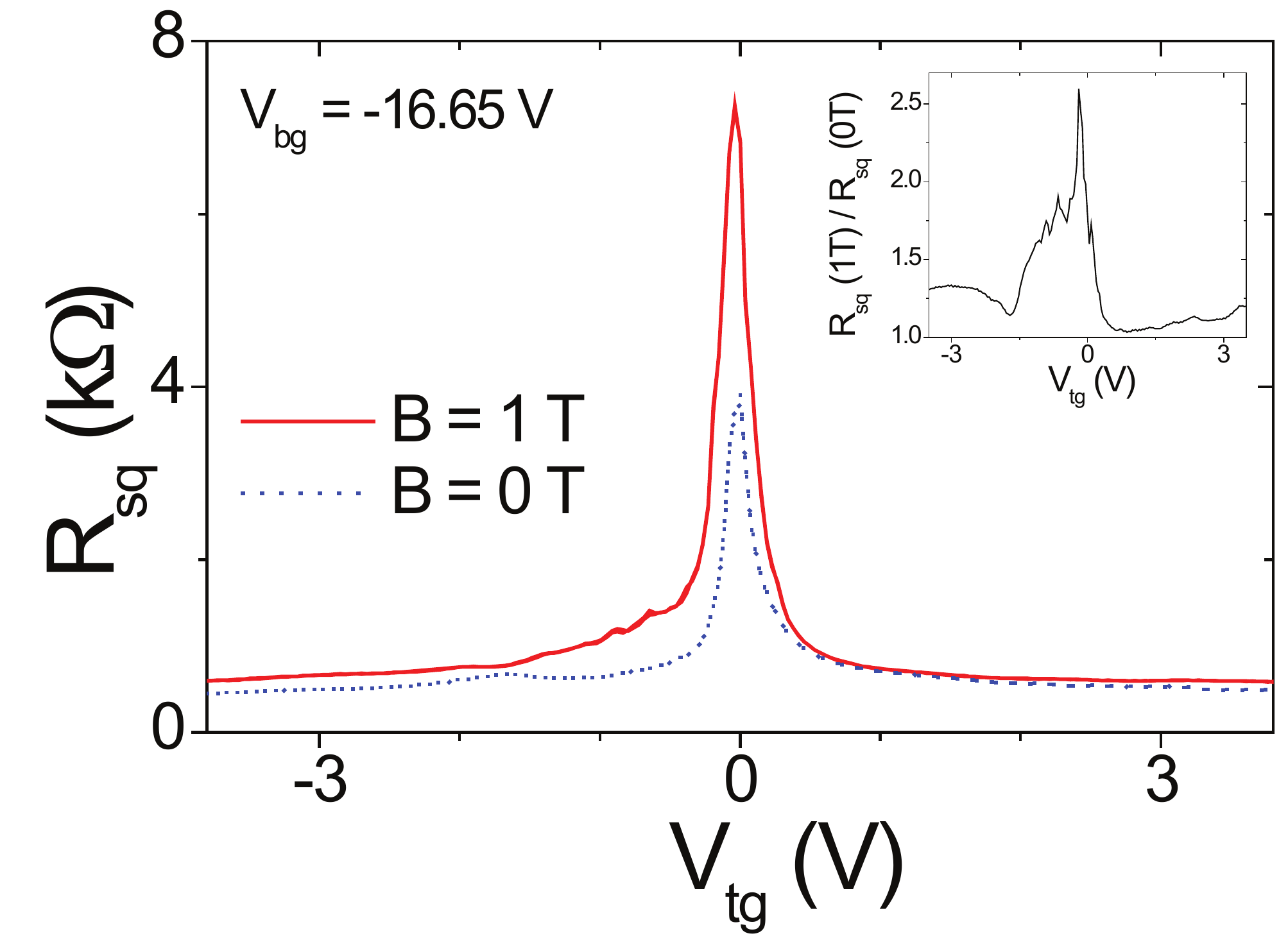}
	\caption{Graphene square resistance as a function of $V_{tg}$ for $V_{bg}=-16.65$ V for sample 1 at 4.2 K and B = 0 (dashed blue line) and 1 T (red solid line). \textit{Inset:} Ratio between the two curves in the main graph showing the magneto-resistance in the sample.}
	\label{fig:SI-magres}
\end{figure}

To be sure that the magneto-resistance contribution is minimal in our nonlocal signals we characterize the graphene sheet resistance as a function of the applied magnetic field.
In Fig. \ref{fig:SI-magres} we show a typical curve of the $R_{sq}$ versus $V_{tg}$ for B = 0 and 1 T measured in a 4-probe geometry.
It can be seen that the change in $R_{sq}$ with $B$ is maximum around the charge neutrality point and small at high values of $n$.
This observation also reflects on our nonlocal measurements in which we observe the monotonic increase in the nonlocal resistance as a function of $B$ characteristic to magneto-resistance effects only for low $n$.

\subsection{Simulation of Hanle precession data for non-homogeneous systems}

\begin{figure}[h]
	\centering
		\includegraphics[width=0.39\textwidth]{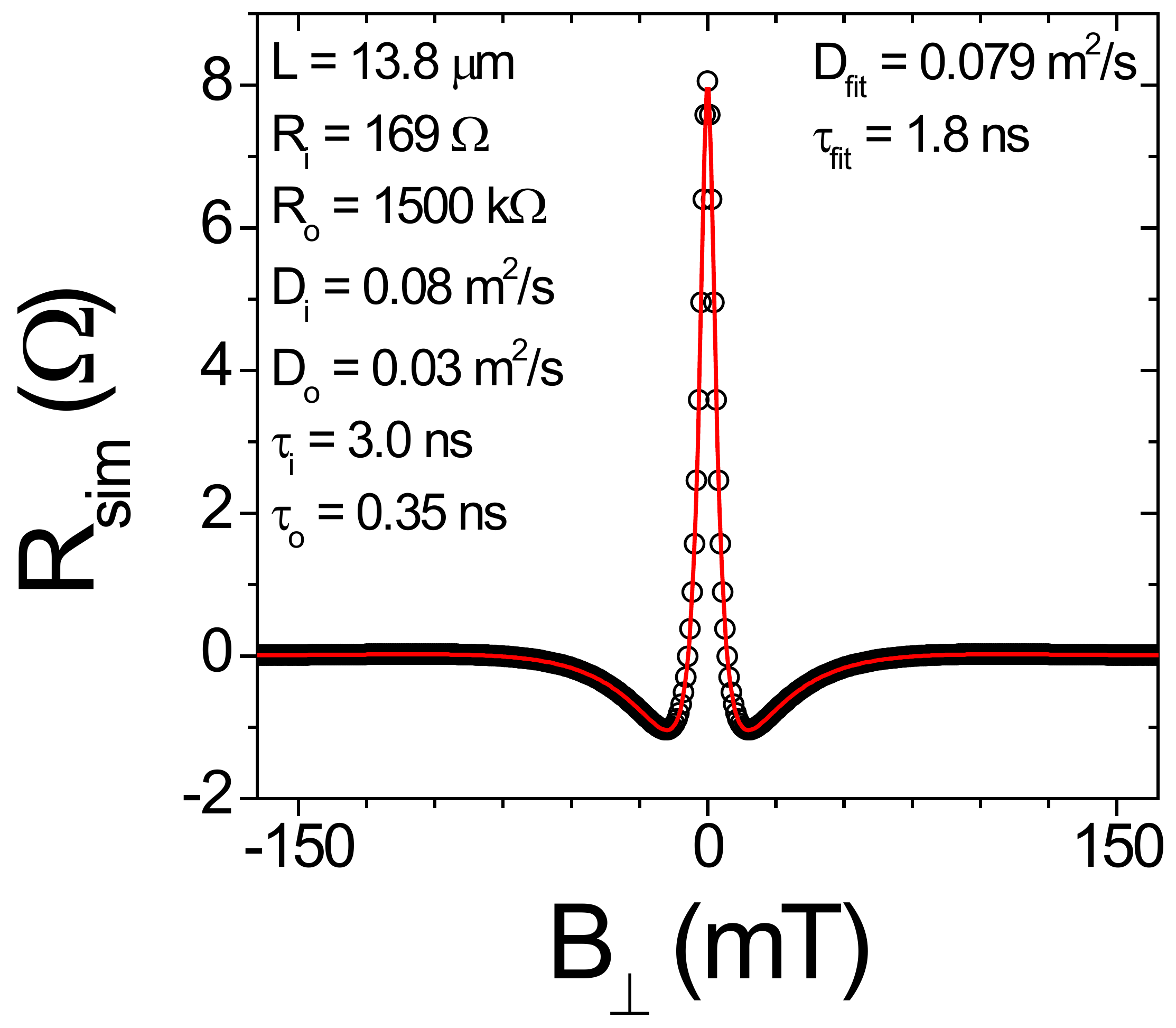}
	\caption{Hanle precession curve calculated using the model for a non-homonegeous system (black dots) with a fit using the expression for the solution of the Bloch equation in a homogeneous system.}
	\label{fig:SI-simhanle}
\end{figure}

In order to describe the different regions in our system (encapsulated and non-encapsulated) we use a model for spin transport in an inhomogeneous system described elsewhere \cite{GuimaraesNanoLetters2012}.
We solve the Bloch equations for spin diffusion in one dimension for a system with three distinct regions connected to one another.
Each one of these regions can have independent values for the relevant parameters.
We assume the outer regions identical, with square resistance $R_{o}$, spin diffusion coefficient $D_{o}$ and spin relaxation time $\tau_{o}$.
The inner region has distinct parameters $R_{i}$, $D_{i}$ and $\tau_{i}$.
We assume that the spins are injected at the left boundary of the inner region and detected on the right, which results in a nonlocal spin signal $R_{sim}$.
For this work we get $R_{o}$, $D_{o}$, $\tau_{o}$, $R_{i}$ and $D_{i}$ by charge and spin transport experiments as described in the main text.
We then fit the simulated curves for $R_{sim}$ using the solution for the Bloch equation for homogeneous systems (exactly in the same way we analyze our experimental results).
From this fit we extract an effective spin diffusion coefficient $D_{fit}$ and relaxation time $\tau_{fit}$.
An example of such simulated curve and fitting is shown in Fig. \ref{fig:SI-simhanle}.

For a fixed $V_{bg}$, the effect of $V_{tg}$ is to change the spin resistance ratio between the inner and outer regions, $\frac{R_{i}\lambda_{i}}{R_{o}\lambda_{o}}$, where $\lambda_{i(o)}$ is the spin relaxation length for the inner (outer) regions.
The change of this ratio affects strongly the extracted spin relaxation times $\tau_{fit}$ but leaves $D_{fit}$ mostly unaffected \cite{GuimaraesNanoLetters2012}.
In order to get a good estimation of $\tau_{i}$, we take the experimental values for $V_{bg}$=-52.5 V and simulate a few sets of points with only changing $R_{i}$.
As it can be seen in Fig. 3b in the main text, we find a close match between our simulated values of $\tau_{fit}$ to our experimentally obtained values of $\tau_{s}$ for $\tau_{i}$ ranging from 3 to 5 ns.
Meaning that the spin relaxation time for the encapsulated region is higher than, but still within the same order of magnitude of, the experimentally obtained values using the solution for the Bloch equations in a homogeneous system.

It is known that $R_{c}$ can affect the measurement of $\tau_{s}$, especially in the case of very long values of $\lambda_{s}$.
However, even though we have a wide range of contact resistances $R_{c}$ 2 - 100 k$\Omega$ with one sample having all contact resistances above 50 k$\Omega$, we did not see any clear correlation between $R_{c}$ and the measured $\tau_{s}$.
We would expect that, in the case of contact induced spin relaxation or poor graphene quality in the regions underneath the contacts, $\tau_{s}$ would be mostly affected in the non-encapsulated regions.
Therefore, the measured spin relaxation time for the non-encapsulated will be an effective spin relaxation time that includes all these effects, and consequently is taken into account in our model by setting $\tau_{o}$ as the measured spin relaxation time for these regions.

\subsection{Experimental Hanle precession data}

\begin{figure}[h!]
	\centering
		\includegraphics[width=0.39\textwidth]{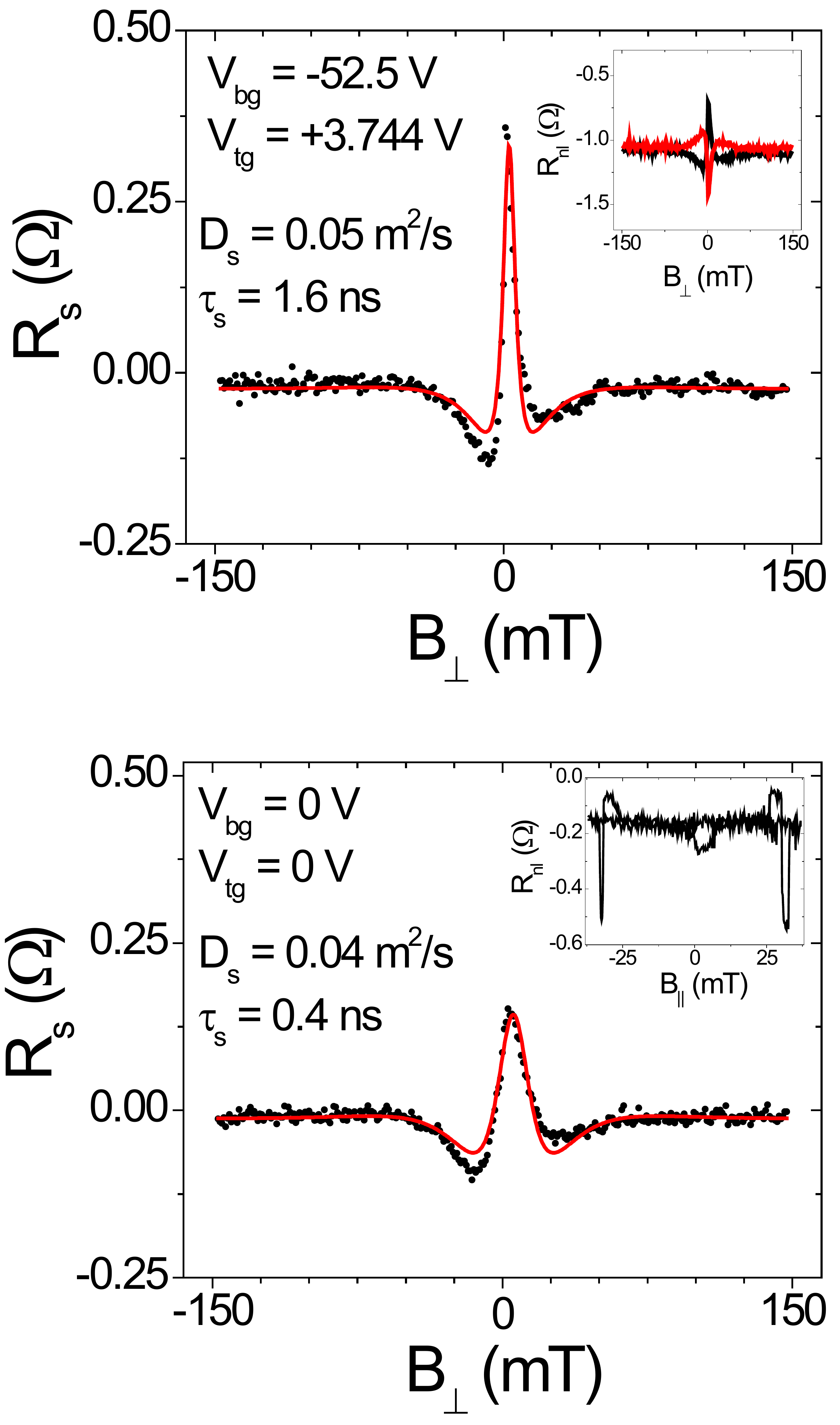}
	\caption{Top: Experimental data for $R_{s}=(R_{nl}^{P}-R_{nl}^{AP})/2$ at $V_{bg}$ = -52.5V and $V_{tg}$ = +3.744 V, where $R_{nl}^{P(AP)}$ is the non-local resistance obtained for the electrodes in a (anti)parallel magnetization (dots) as a function of an applied perpendicular magnetic field. The fit (red line) used to extract $D_{s}$ and $\tau_{s}$ is also shown. In the inset it is shown the data for $R_{nl}^{P(AP)}$ in black (red). Bottom: Hanle precession curve for $V_{bg}$ = $V_{tg}$ = 0 V with the respective non-local spin-valve shown in the inset.}
	\label{fig:SI-hanle}
\end{figure}

For most of our devices we observe a small asymmetry in our Hanle precession curves between negative and positive perpendicular magnetic fields \ref{fig:SI-hanle} which might arise due to some anisotropy in the magnetization of the electrodes.
Although we cannot be certain about the reason for this asymmetry, the parameters extracted by fitting just one or the other side of the curve do not change significantly.
Comparing the parameters extracted from each side, $\tau_{s}$ deviates by a factor of $\approx$ 1.06 and $D_{s}$ by a maximum of a factor 2.
Since the values for $D_{s}$ extracted by fitting the whole Hanle curve match the values for $D_{c}$ extracted by charge transport measurements, we believe that our procedure is justified.

The measured spin polarization of our electrodes were in the range $P$ = 1 - 6 $\%$.
These values for $P$ are in the lower range of the values reported in literature \cite{MihaiPhys.Rev.B2009,KawakamiPhys.Rev.Lett.2011}, however we believe that this value can be improved by increasing the quality of the high resistive barriers \cite{KawakamiPhys.Rev.Lett.2010}.

\subsection{Results for spin and charge transport at 4.2 K and room temperature}

\begin{figure*}[h]
	\centering
		\includegraphics[width=0.8\textwidth]{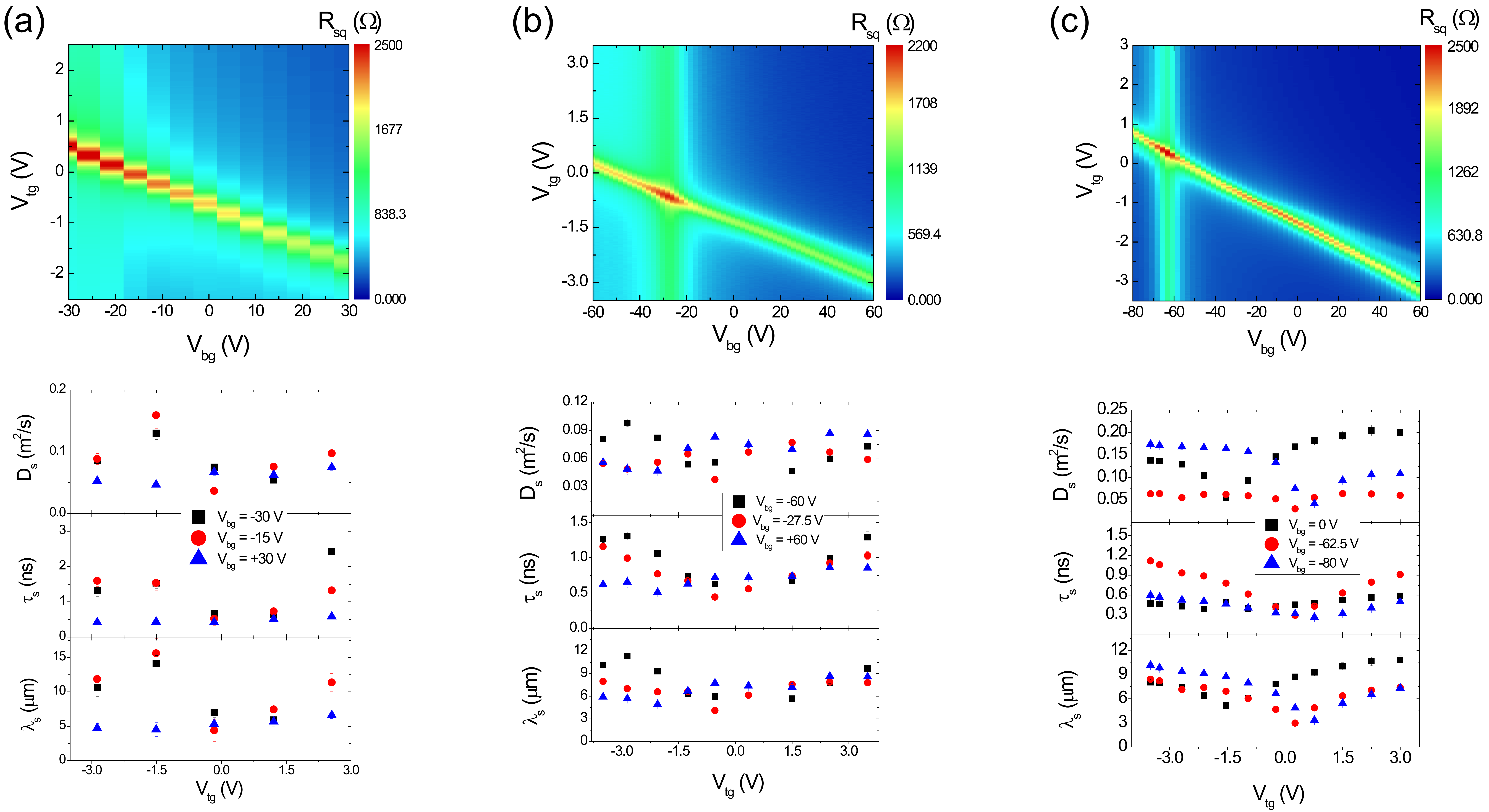}
	\caption{Charge and spin transport measurements for (a) sample 1 at room temperature and sample 2 at (b) room temperature and (c) 4.2 K.}
	\label{fig:samples}
\end{figure*}

Our results presented in the main text were all performed in sample 1 at 4.2 K.
For comparison we also include here the square resistance as a function of $V_{tg}$ and $V_{bg}$ and the extracted spin transport parameters for the same sample at room temperature (Fig. \ref{fig:samples}a) and another sample (sample 2) at room (Fig. \ref{fig:samples}b) and liquid Helium (Fig. \ref{fig:samples}c) temperatures.
The samples dimensions and their electronic mobilities at room temperature and 4.2 K are given in table \ref{fig:samples}.
Sample 3 (not shown here) showed similar results.

\begin{table*}[h]
\centering
\begin{tabular}{|c|c|c|c|c|c|c|}
\hline
 & $\ell_{tg}$ & $\ell_{enc}$ & $L$ & $W$ & $\mu_{RT}$ & $\mu_{4K}$ \\
\hline
Sample 1 & 10.35 $\mu$m & 12 $\mu$m & 13.8 $\mu$m & 1.3 $\mu$m & 1.5 m$^{2}$/Vs & 2.3 m$^{2}$/Vs \\
\hline
Sample 2 & 4.5 $\mu$m & 7.6 $\mu$m & 8.72 $\mu$m & 2.8 $\mu$m & 2.4 m$^{2}$/Vs & 3.3 m$^{2}$/Vs \\
\hline
\end{tabular}
\label{tab:samples}
\caption{Length of the top gated region ($\ell_{tg}$), encapsulated region ($\ell_{enc}$), inner electrodes spacing ($L$), graphene flake width ($W$) and electronic mobilities at room temperature ($\mu_{RT}$) and 4.2 K ($\mu_{4K}$) for samples 1 and 2.}
\end{table*}

\subsection{Estimation of $\Delta_{R}$}

In order to give a rough estimation for the spin orbit coupling strength $\Delta_{R}$ we assume that the spin relaxation is composed of the sum of an electric field independent and electric field dependent terms: $1 / \tau_{\bot} = 1 / \tau_{ind} + 1 / \tau_{E}$.
The spin relaxation time $\tau_{||}$ is taken equal to the spin relaxation experimentally ($\tau_{s}$) and $\tau_{ind}$ is $\tau_{s}$ multiplied by the ratio $r = \tau_{\bot} / \tau_{s}$ at \linebreak $\bar{E}$ = 0 V/nm: $\tau_{ind} = 0.75 \tau_{s}$.
If we assume a D'Yakonov-Perel spin relaxation mechanism for $1 / \tau_{E} = \frac{4 \Delta_{R}^2}{\hbar^{2}} \tau_{p}$, we have: $\Delta_{R}=\zeta \bar{E}=(\hbar / 2) (\tau_{E} \tau_{p})^{-1/2}$ \cite{FabianPhys.Rev.B2009a}, where $\tau_{p}=2 D_{c} / v_{F}$ is the momentum relaxation time, $D_{c}$ the charge diffusion constant and $v_{F}\approx 10^{6}$ m/s the Fermi velocity.
We further take $D_{c} \approx D_{s}$.
Solving it for $\tau_{E}$ we have:

\begin{equation}
\tau_{E} = \tau_{s} \left( \frac{1}{r} - \frac{1}{0.75} \right)^{-1},
\end{equation}

and

\begin{equation}
\zeta = \frac{\Delta_{R}}{\bar{E}} = \frac{\hbar}{2 \bar{E}} \frac{1}{\sqrt{\tau_{s} \tau_{p}}} \left( \frac{1}{r} - \frac{1}{0.75} \right)^{-1/2}
\end{equation}

\section{D'Yakonov-Perel \textit{versus} Elliott-Yafet mechanisms for spin relaxation}
Two mechanisms are often considered in order to try to explain spin relaxation in graphene: they are the D'Yakonov-Perel (DP) and Elliott-Yafet (EY) mechanisms for spin relaxation.
These two mechanisms show different behavior of the spin relaxation time as a function of the momentum relaxation time ($\tau_{p}$): $\tau_{EY} \propto \tau_{p}$ and $\tau_{DP} \propto 1 / \tau_{p}$, with $\tau_{EY(DP)}$ been the spin relaxation time due to the EY (DP) mechanism.

It was theoretically demonstrated that $\tau_{EY}$ for the specific case of graphene for many different type of scatterers is given by \cite{GuineaPhys.Rev.Lett.2012}: 

\begin{equation}
\tau_{EY} \approx \frac{E_{F}^{2}}{\Delta_{SO}^2} \tau_{p},
\end{equation}

\noindent where $E_{F}$ is the Fermi energy and $\Delta_{SO}$ the spin orbit coupling.
The DP mechanism obeys:

\begin{equation}
\tau_{DP} \approx \frac{\hbar^{2}}{\Delta_{SO}^2} \frac{1}{\tau_{p}}.
\end{equation}

Following the reasoning of Huertas-Hernando et al. \cite{BrataasPhys.Rev.Lett.2009}, to obtain the relative importance of these mechanisms we take the ratio:

\begin{equation}
\frac{\tau_{EY}}{\tau_{DP}} \approx \left( \frac{E_{F} \tau_{p}}{\hbar} \right)^{2}.
\end{equation}

If $\tau_{EY} / \tau_{DP} >$ 1, the spin relaxation time due to the DP mechanism is smaller and it dominates over the EY mechanism.
Using typical values for our samples of $E_{F} \approx$ 30 meV (equivalent to a charge carrier density $n \approx$ 10$^{12}$ cm$^{-2}$) and $\tau_{p} \approx$ 10$^{-13}$ s, we obtain $\tau_{EY} / \tau_{DP} \approx$ 20, indicating a higher importance of the DP mechanism for the spin relaxation in our samples.

Although this is not a definite proof that the DP is the dominant mechanism for spin relaxation in graphene, we believe that it is the most appropriate way to obtain the spin orbit coupling strength due to the electric field in our devices.


%

\end{document}